\documentclass{article}
\usepackage{amssymb}
\usepackage{amsmath}
\usepackage{amsfonts}
\usepackage{amsthm}
\usepackage[T1]{fontenc}
\usepackage{authblk}

\newtheorem{remark}{Remark}
\newcommand*{\bfrac}[2]{\genfrac{}{}{0pt}{}{#1}{#2}}

\begin{document}

\title{Special Killing forms on toric Sasaki-Einstein manifolds}

\author[1]{Vladimir Slesar\thanks{vlslesar@central.ucv.ro}}
\author[2]{Mihai Visinescu\thanks{mvisin@theory.nipne.ro}}
\author[3,4]{Gabriel Eduard
V\^ilcu\thanks{gvilcu@upg-ploiesti.ro}\thanks{gvilcu@gta.math.unibuc.ro}}
\affil[1]{Department of Mathematics, University of Craiova,

Str. Al.I. Cuza, Nr. 13, Craiova 200585, Romania}
\affil[2]{Department of Theoretical Physics,

National Institute for Physics and Nuclear Engineering,

Magurele, P.O.Box M.G.-6, Romania}
\affil[3]{Department of Mathematical Modelling,

Economic Analysis and Statistics,

Petroleum-Gas University of Ploie\c sti,

Bulevardul Bucure\c sti, Nr. 39, Ploie\c sti, 100680, Romania}
\affil[4]{Faculty of Mathematics and Computer Science,

Research Center in Geometry, Topology and Algebra,

University of Bucharest, Str. Academiei, Nr. 14, Sector 1,
Bucharest, 060042, Romania}
\date{\today }
\maketitle

\begin{abstract}
In this paper we study the interplay between complex coordinates on the
Calabi-Yau metric cone and the special Killing forms on the toric
Sasaki-Einstein manifold. In the general case we give a procedure to locally
construct the special Killing forms. In the final part we exemplify the
general scheme in the case of the $5-$dimensional $Y^{p,q}$ spaces, identifying the
additional special Killing 2-forms which were previously obtained by the second author of the present paper, but with a different method, in [Mod. Phys. Lett. A \textbf{27}
(2012) 1250217].

\end{abstract}



\section{Introduction}

Historically, Sasakian structures grew out of research in contact
geometry and were studied extensively in the 1960's. In the last time,
there has been renewed interest in Sasaki geometries in connection with
some recent developments in mathematics and theoretical physics \cite{JS}.
New explicit inhomogeneous Sasaki-Einstein metrics in all dimensions have
been constructed in \cite{GMSW2}. Moreover, using the construction of
Boyer and Galicki \cite{BG2}, infinitely many toric Sasaki-Einstein
manifolds with arbitrarily high second Betti number of every dimension
$\geq5$ were given in \cite{CvC}.

A particularly interesting class of toric contact structures on $S^2
\times S^3$ have been studied by physicists \cite{GMSW,M-S} and denoted
by $Y^{p,q}$ where $p, q$ are relative prime integers satisfying $0< q
<p$. These structures have become of much interest in connection with
the AdS/CFT conjecture \cite{JMM}. The AdS/CFT correspondence provides
a detailed correspondence between certain conformal field theories and
geometries leading to remarkable new results in both sides. The
Sasaki-Einstein manifolds with their explicit metrics offer useful
models to test AdS/CFT duality.

The purpose of this paper is the explicit construction of the special
Killing forms on toric Sasaki-Einstein manifolds, this goal being achieved
using foliated coordinates on the metric cone of the manifold. The interest
in the Killing forms is motivated by the key role of symmetries in physics.
For the geodesic motions in a space-time, the usual conserved quantities
are related to the isometries which correspond to Killing vectors.
Sometimes a space-time could admit higher order symmetries described by
symmetric St\"{a}ckel-Killing tensors. These symmetries are known as
{\it hidden symmetries}, and the corresponding conserved quantities are
quadratic or, more general, polynomial in momenta.

Another natural generalization of the Killing vector fields is represented by
the Killing forms which in many aspects are more important than the
St\"{a}ckel-Killing tensors. It was realized the significant connection
between Killing forms and {\it nonstandard supersymmetries}.
Investigation of the properties of higher dimensional space-times has
pointed out the role of conformal Killing forms to generate black holes
solutions \cite{FK}.

In the case of Sasaki-Einstein spaces, there are special Killing forms directly
constructed from the contact form of the Sasaki manifolds. Besides
these special Killing forms, there are two special Killing forms connected with the
additional parallel forms of the Calabi-Yau cone manifolds given by the
complex volume and its conjugate \cite{US}.

Concerning special Killing forms, we would like to point out that this notion was originally introduced by Tachibana and Yu in \cite{TY}, but the concept was mainly developed by Semmelman \cite{US}, who gave some equivalent definitions more natural than the original one, and a complete description of manifolds admitting special Killing forms. It is easy to remark that the restriction from Killing forms to special Killing forms is analogous to the definition of a Sasakian structure as a unit Killing vector field satisfying an additional equation. We also note that special Killing forms are of great interest in physics, due to potential applications to gauge/gravity duality and string theory. For example, very recently, Houri, Takeuchi and Yasui \cite{HTY} introduced the concept of a \emph{Sasaki manifold with torsion} and proved that these metrics admit two types of hidden symmetries, the first arising from the special Killing forms. Moreover, they showed that these metrics on the five-dimensional compact manifolds naturally generalize the toric Sasaki-Einstein metrics $Y^{p,q}$ and $L^{abc}$.

The great importance of special Killing forms in our setting comes also from the fact that any Killing form of degree at least 3 on a compact Sasaki-Einstein manifold is actually special \cite{STRO}. In particular, since 3-Sasakian manifolds are automatically Einstein \cite{KSW}, the above result extends on compact 3-Sasakian spaces, an interesting class of manifolds investigated in the context of the AdS/CFT-correspondence by Yee \cite{YEE}, Martelli, Sparks and Yau \cite{MSY}.

The organization of the paper is as follows: In the next Section we
give some mathematical preliminaries regarding the special Killing
forms on Sasaki-Einstein manifolds and their Calabi-Yau cones. In order
to construct the special Killing forms on Sasaki-Einstein manifolds, in
Section 3 we investigate the complex volume and the foliated
coordinates on the metric cone, and then we give a procedure to locally extract the special Killing forms on a toric Sasaki-Einstein manifold. In Section 4 we exemplify the general scheme in the case of the $Y^{p,q}$ manifolds, identifying two additional special Killing 2-forms. We note that these Killing forms were also obtained in [22] with a different approach, namely by a direct calculation. Finally we give our conclusions in Section 5.

\section{Preliminaries}

\label{geometric objects}

The conformal Killing-Yano tensors provide a generalization of
Killing vector fields. In order to introduce this concept, we start out by
considering the tensor product space $V^{*}\otimes \bigwedge^pV^{*}$, where
$\left( V,\left\langle \cdot ,\cdot \right\rangle \right) $ is an $m-$
dimensional Euclidean space. Considering the wedge product $\wedge $ and the
contraction map $\lrcorner $, this tensor product splits as a direct sum of three terms,
with the following identification (see \cite{US})
\[
V^{*}\otimes \bigwedge^pV^{*}\simeq \bigwedge^{p+1}V^{*}\oplus
\bigwedge^{p-1}V^{*}\oplus \bigwedge^{p,1}V^{*} \,.
\]
Let us denote by $\mathrm{pr}_{\bigwedge^{p,1}}$, $\mathrm{pr}_{\bigwedge^{p+1}}$,
$\mathrm{pr}_{\bigwedge^{p-1}}$ the
projections on the first, second and third term, respectively.

The following obvious relation is satisfied
\[
\mathrm{pr}_{\bigwedge^{p,1}}=\mathrm{Id}-\mathrm{pr}_{\bigwedge^{p+1}}-\mathrm{pr}_{\bigwedge^{p-1}}
\,,
\]
so if $\{e_i\}_{1\le i\le m}$ is an orthonormal basis and
$\{\alpha^i\}_{1\le i\le m}$ the associated dual basis, then the above
projection can be written as
\begin{eqnarray}
\mathrm{pr}_{\bigwedge^{p,1}}\left( \alpha \otimes \Psi \right)  &=&\alpha \otimes
\Psi -\frac 1{p+1}\sum_i\alpha ^i\otimes e_i\lrcorner \left( \alpha \wedge
\Psi \right)   \label{proiectie} \\
&&-\frac 1{n-p+1}\sum_i\alpha ^i\otimes \alpha ^i\wedge \left( \alpha^{\sharp }
\lrcorner \Psi \right) \,.  \nonumber
\end{eqnarray}
for any $\alpha \in V^{*}$ and $\Psi \in \wedge ^{p} V^{*}$, $\sharp $ being the
"musical" isomorphism between $V^{*}$ and $V$, related to the Euclidean
structure. We recall that given an element $\alpha\in V^{*}$, we know that there exists a unique $v\in V$  such that $\alpha(u)=\langle v,u\rangle$  for all $u\in V$  and the "sharp" operator $\sharp$ is defined by $\alpha^{\sharp}=v$.

It turns out that $\bigwedge^{p,1}V^{*}$ is the intersection of the
kernels of wedge product and contraction map. This fact can be also checked
directly, by considering
\begin{eqnarray*}
\mu _1 :V^{*}\otimes \bigwedge^p V^{*}\rightarrow
\bigwedge^{p+1}V^{*},\,\mu _1\left( \alpha \otimes \Psi \right) :=\alpha
\wedge \Psi \,, \\
\mu _2 :V^{*}\otimes \bigwedge^p V^{*}\rightarrow
\bigwedge^{p-1}V^{*},\,\mu _2\left( \alpha \otimes \Psi \right)
:=\alpha^{\sharp }
\lrcorner \Psi \,,
\end{eqnarray*}
and calculating
\[
\mu _1\circ \mathrm{pr}_{\bigwedge^{p,1}}=0,\,\,\mu _2\circ
\mathrm{pr}_{\bigwedge^{p,1}}=0\,.
\]

Now, let us consider a Riemannian manifold $\left( M,g\right) $ with
$\mathrm{dim} M=m$
endowed with the Levi-Civita connection $\nabla $. If $\Psi \in \Gamma
\left(\bigwedge^pTM^{*}\right)$ is a smooth
differential form of dimension $p$, then we can regard $\nabla \Psi $ as a
section in a tensorial fibre bundle, $\nabla \Psi \in \Gamma \left(TM^{*}\right)\otimes \Gamma
\left(\bigwedge^pTM^{*}\right)$, and we have
\begin{equation}
\nabla \Psi =\sum_i\alpha ^i\otimes \nabla _{e_i}\Psi \,,
\label{tensor connection}
\end{equation}
where $\{e_i\}_{1\le i\le m}$ and $\{\alpha^i\}_{1\le i\le m}$ are now the
corresponding local orthonormal frames.

The \emph{twistor operator} $T$ is defined as the composition \cite{US}
\[
T:=\mathrm{pr}_{\bigwedge^{p,1}}\circ \nabla\,,
\]
and using (\ref{proiectie}) and (\ref{tensor connection}), we obtain
\begin{eqnarray*}
T\left( \Psi \right)&:=&\mathrm{pr}_{\bigwedge^{p,1}}\left( \nabla \Psi \right) \\
\ &=&\sum_i\alpha ^i\otimes \nabla _{e_i}\Psi
-\frac 1{p+1}\sum_i\alpha^i\otimes e_i\lrcorner \left( \sum_j\alpha ^j\wedge
\nabla _{e_j}\Psi \right)
\\
&&\ +\frac 1{n-p+1}\sum_i\alpha ^i\otimes \alpha ^i\wedge \left(
-\sum_je_j\lrcorner \nabla _{ej}\Psi \right) \\
\ &=&\sum_i\alpha ^i\otimes \left( \nabla _{e_i}\Psi -\frac
1{p+1}e_i\lrcorner d\Psi +\frac 1{n-p+1}\alpha ^i\wedge d^{*}\Psi \right) \,,
\end{eqnarray*}
where $d$ and $d^{*}$ are the de Rham differential and co-differential
operators with respect to the Riemannian structure $g$.

It is now obvious that $T\left( \Psi \right) =0$ if and only if
\[
\nabla _X\Psi -\frac 1{p+1}X\lrcorner d\Psi +\frac 1{n-p+1}X^{*}\wedge
d^{*}\Psi =0\,,
\]
for any vector field $X$. The differential forms that satisfy the above
equation are called \emph{conformal Killing -Yano tensors} or \emph{conformal
Killing forms} \cite{US}.

Coclosed conformal Killing form are called \emph{Killing forms}. If the
dimension is $1$, then Killing forms are just dual to Killing vector fields.

Any parallel form is a Killing form. It is easy also to see that on
$\left(\mathbb{R}^3,x^1,x^2,x^3\right) $ with canonical metric, the
differential form
\[
\Psi =x^3dx^1\wedge dx^2+x^1dx^2\wedge dx^3+x^2dx^3\wedge dx^1
\]
is a Killing form. Other non-trivial Killing and conformal Killing form can
be defined on standard sphere, on Sasakian manifolds, on nearly K\"ahler
manifolds.

A particular class of Killing forms is represented by the
\emph{special Killing forms}, which satisfy, for some constant $c$, the equation
\cite{TY,US}
\[
\nabla _X\left( d\Psi \right) =cX^{*}\wedge \Psi \,.
\]
Here $X$ is an arbitrary vector field on $M$. The interesting feature of special
Killing forms comes from the fact that the most known Killing forms are
actually special.

In order to classify this type of differential forms, in \cite{US}
Semmelmann introduced a correspondence between special Killing forms defined
on the manifold $M$ and parallel forms defined on the metric cone $C(M)$.
This is in fact the product manifold $M\times \mathbb{R}_{>0}$, with
$\dim C(M)=2n=m+1$, endowed with the warped metric
$g_{cone}:=dr^2+r^2g$. More exactly, a $p-$dimensional differential form
$\Psi $ is a special Killing form on $M$ if and only if the corresponding
form
\begin{equation}
\Psi _{cone}:=r^pdr\wedge \Psi +\frac{r^{p+1}}{p+1}d\Psi
\label{eq Semmelmann}
\end{equation}
is parallel on $C(M)$.

A particular example when the special Killing forms are known to exist is
represented by Sasaki-Einstein manifolds. It is well-known that a Sasakian
manifold is a Riemannian manifold $(M,g)$ whose metric cone $C(M)$ is
K\"{a}hler \cite{BG}. From this it follows that $M$ also has a contact
structure $(\phi,\xi,\eta)$ such that the 1-dimensional foliation generated
by the Reeb vector field $\xi$ is transversely K\"{a}hler \cite{FOW}.
A Sasaki-Einstein manifold is a Riemannian manifold $(M,g)$  that is both
Sasakian and Einstein. On the other hand, a toric Sasakian manifold is a
Sasakian manifold $(M,g)$ of dimension $2n-1$ with Sasakian structure
$(\phi,\xi,\eta)$,  such that there is an effective action of
$n$-dimensional torus $G$, preserving the Sasakian structure, while
$\xi$ is an element of the Lie algebra of $G$ \cite{CFO}. Equivalently, a
toric Sasakian manifold is a Sasakian manifold
whose K\"{a}hler cone is a toric K\"{a}hler manifold \cite{VG1,VG2}.

On a $(2n-1)-$dimensional Sasaki manifold with the Reeb vector field
$\xi$ and $1-$form $\eta := \xi^\ast$, there are the following special
Killing forms:
\[
\Psi_k = \eta \wedge (d\eta)^k \quad , \quad k = 0, 1,
\cdots , n-1\,.
\]
Besides these Killing forms, there are $n-1$ closed conformal Killing
forms (also called $\ast$-Killing forms)
\[
\Phi_k = (d\eta)^k \quad , \quad k = 1, \cdots , n-1 \,.
\]

In the case of a Sasaki-Einstein manifold the metric cone is
Ricci-flat, has holonomy $SU_n$ and there are two additional parallel forms
related to complex volume form. To wit, on this type of manifolds with
$SU_n$ geometric structure, there exists a complex volume form $\Omega $ and
also its conjugate $\bar \Omega $, these globally defined complex $n-$dimensional
differential forms being furthermore parallel \cite{Mor,US,VV}. Then,
special real parallel differential forms on the Sasaki-Einstein metric can
be derived.

From the above consideration, it becomes clear the relevance of the complex
volume form of the metric cone of a Sasaki-Einstein manifold. Consequently,
it is of interest to consider in the following the complex coordinates defined on a toric
Sasaki-Einstein manifold. The geometric cone of this manifold can be regarded
in the classical manner
\cite{Abr,M-S,M-S-Y}.

First of all, we introduce symplectic (action-angle) coordinates
$(y^i,\phi ^i)$; the angular coordinates $\phi ^i$ will generate the
toric action. The corresponding K\"ahler metric on $C(M)$ will be \cite{M-S-Y}
\[
ds^2=G_{ij}dy^idy^j+G^{ij}d\phi ^id\phi ^j\,,
\]
where the metric coefficients are obtained using the symplectic potential $G$
\[
G_{ij}=\frac{\partial ^2G}{\partial y^i\partial y^j},
\]
and $\left(G^{ij}\right)=\left( G_{ij}\right) ^{-1}$.

We can now express the complex structure
\[
J=\left(
\begin{array}{cc}
0 & -G^{ij} \\
G_{ij} & 0
\end{array}
\right) \,,
\]
while the K\"ahler form is
\[
\omega =dy^i\wedge d\phi ^i\,.
\]

Next, we present the complex approach. From symplectic coordinate we can pass
to the coordinate patch $(x^i,\phi ^i)$, obtained from complex coordinates
$z^i:=x^i+\mathrm{i}\phi ^i$, with $\mathrm{i}:=\sqrt{-1}$. This time, the
metric structure is written in the following manner
\[
ds^2=F_{ij}dx^idx^j+F_{ij}d\phi ^id\phi ^j\,.
\]
In this setting, the metric coefficients are again obtain using the Hessian
of the K\"ahler potential $F$, i.e.
\[
F_{ij}=\frac{\partial ^2F}{\partial x^i\partial x^j}\,.
\]

With respect to the coordinates $(x^i,\phi ^i)$, the K\"ahler form is
\[
\omega =\left(
\begin{array}{cc}
0 & F_{ij} \\
-F_{ij} & 0
\end{array}
\right) \,.
\]
The symplectic potential $G$ and the K\"ahler potential $F$ are related by the
Legendre transform
\[
F(x)=\left( y^i\frac{\partial G}{\partial y^i}-g\right) \:\left( y=\partial
F/\partial x\right) \,.
\]
Consequently, $F$ and $G$ are Legendre dual to each other
\[
F(x)+G(y)=\sum_j\frac{\partial F}{\partial x^j}\frac{\partial G}
{\partial y^i}\,\,\mbox{at}\,x^i=\frac{\partial G}{\partial y^i}\,\,
\mbox{or}\,y^i=\frac{\partial F}{\partial x^i} \,.
\]
From the above considerations it follows that $F_{ij}=$ $G^{ij}$
($y=\partial F/\partial x$).

Our interest for complex coordinates on Sasakian toric manifolds comes from
the fact that using this particular type of coordinates it is possible to
express a complex volume form in a very convenient way. More exactly, on
these manifolds the complex volume form $\Omega $ can be
written as \cite{M-S-Y}
\[
\Omega =e^{z^1}dz^1\wedge ..\wedge dz^n \,.
\]
Employing the above relation, in the next sections we show that is possible
to extract the special Killing forms on manifolds of Sasaki-Einstein type.

\section{Complex volume form and foliated coordinates on geometric cone}

Within this section, we show that working with foliated coordinates we can
locally extract the special Killing forms on a Sasaki-Einstein manifold.

More precisely, starting with the natural foliated structure on the metric
cone $C(M)$ (which topologically can be identified with
$\mathbb{R}_{>0}\times M$), we consider in the following the existence of (local) foliated
coordinates $\left( r;f^2,..,f^n,\phi ^1,..,\phi ^n\right) $. $r$ will stands
for the \emph{transverse coordinate}, while $f^2,..,f^n,\phi ^1,..,\phi ^n$ will
stand for \emph{leafwise coordinates}, describing the immersed submanifolds; for general
topics concerning foliated structures we indicate \cite{To}. As the coordinates $(x^i,\phi ^i)$ are assumed independent,
the smooth link between the initial local coordinates
$\left( x^i,\phi ^i\right) $
and the foliated coordinates $\left( r;f^i,\phi ^i\right) $ is expressed in
the form

\[
\left\{
\begin{array}{l}
x^i=x^i\left( r,f^2,..,f^n\right) \,, \\
\phi ^i=\phi ^i \,.
\end{array}
\right.
\]
Consequently, for the coframes $\left( dx^i,d\phi ^i\right) $ and $\left(
dr;df^i,d\phi ^i\right) $ we get
\[
\left\{
\begin{array}{l}
dx^i=\frac{\partial x^i}{\partial r}dr+\frac{\partial x^i}{\partial f^j}df^j
\,,
\\
d\phi ^i=d\phi ^i \,.
\end{array}
\right.
\]

We consider also the Jacobi matrix
\[
\mathcal{A}:=\left(
\begin{array}{llll}
\frac{\partial x^1}{\partial r} & \frac{\partial x^1}{\partial f^2} & \cdots
& \frac{\partial x^1}{\partial f^n} \\
\vdots & \vdots & \vdots & \vdots \\
\frac{\partial x^n}{\partial r} & \frac{\partial x^n}{\partial f^2} & \cdots
& \frac{\partial x^n}{\partial f^n}
\end{array}
\right) \,.
\]

In accordance with (\ref{eq Semmelmann}) (see also \cite{US,Vis}), we search
for a complex differential form $\omega ^M$ which verifies the relation
\begin{equation}
\Omega =r^{n-1}dr\wedge \omega ^M + \frac{r^{n}}{n}d\omega ^M \,.
\label{ecuatie Sem}
\end{equation}

The starting point is the following description of the complex volume form
$\Omega $ using foliated coordinates
\begin{eqnarray*}
\Omega &=&e^{z^1}\left( \frac{\partial x^1}{\partial r}dr+\frac{\partial x^1}
{\partial f^j}df^j+\mathrm{i}d\phi ^1\right) \wedge \dots \\
&&\dots \wedge \left( \frac{\partial x^n}{\partial r}dr+\frac{\partial x^n}
{\partial f^j}df^j+\mathrm{i}d\phi ^n\right) \,.
\end{eqnarray*}
In order to extract $\omega ^M$, we need to keep the trace of the
differential form $dr$. We get $n$ terms in the above wedge product;
considering $\Omega $ as a sum of differential forms of dimension $n$, for
an arbitrary term let us assume that $i$ is the rank of $dr$, and in the
left we have the ranks $j_1$,..,$j_p$ where terms of type $\mathrm{i}d\phi ^s$
appear in the wedge product, while in the right of $dr$ we get also the ranks
$k_1$,..,$k_q$ where terms of type $\mathrm{i}d\phi ^s$ also appear. Tracing
out all the terms that appear, we finally get
\begin{equation}
\omega ^M=\frac 1{r^{n-1}}e^{z^1}\sum_{i=1}^n\frac{\partial x^i}{\partial r}
\sum_{\bfrac{p,q}{C_i}}\mathrm{i}^{p+q}\sum_{\bfrac{J,K,L}{C_{i,p,q}}}\left(
-1\right) ^{\mathcal{S}}
\mathcal{A}_{J;i;K}^Ldf^L\wedge d\phi ^J\wedge d\phi ^K \,,  \label{suma}
\end{equation}
where in the above expression we use multi-indices $J:=\left(
j_1,..,j_p\right) $ with $j_1<..<j_p$, $K:=\left( k_1,..,k_q\right) $ with
$k_1<..<k_q$ and $L:=\left( l_1,..,l_{n-p-q+1}\right) $ with
$l_1<..<l_{n-p-q+1}$; the condition $C_i$ means $1\le p\le i-1$, and $i\le
q\le n$, the condition $C_{i,p,q}$ means
$1\le j_1<..<j_p<i$, $i<k_1<..<k_q\le n$
and $2\le l_1<..<l_{n-p-q+1}\le n$. We also denote
\begin{eqnarray}
df^L&:=&df^{l_1}\wedge
\dots \wedge df^{l_{n-p-q+1}},\nonumber\\ d\phi ^J&:=&d\phi ^{j_1}\wedge
\dots \wedge d\phi^{j_p}\,,\nonumber\\
d\phi ^K&:=&d\phi ^{k_1}\wedge \dots \wedge d\phi ^{k_q} \,.\nonumber
\end{eqnarray}
Furthermore, $\mathcal{A}_{J;i;K}^L$ stands for the determinant of the matrix obtained
from $\mathcal{A}$ by suppressing the rows
$j_1$,..,$j_p$, $i$, $k_1$, ..,$k_q$, and selecting the columns $l_1$,..,$l_{n-p-q+1}$; if no indices of a certain
type exist, then we agree to denote this by $0$ . For instance, for the term
$\mathcal{A}_{J;i;K}^0$, which corresponds to the case when the term is
constructed only using wedge product of differential forms of type
$d\phi^{j_s}$ and $d\phi ^{k_s}$, we agree to put $\mathcal{A}_{J;i;K}^0:=1$.
Finally, concerning the sign of each term in the above sum, we have
\begin{eqnarray*}
\mathcal{S} &=&nq-\sum_{s=1}^qk_s+np-\sum_{s=1}^pj_s \\
&&-\frac{q\left( q-1\right) }2-\frac{p\left( p-1\right) }2-p\left(
q+1\right) \,.
\end{eqnarray*}

\begin{remark}
Let us notice that according to \cite[Lemma 4.5]{US},
the differential form $\omega ^M$ obtained from (\ref{ecuatie Sem}) will no
longer depend on the transverse coordinate $r$. Consequently, $\omega ^M$
will be a complex special Killing form on the manifold $M$, locally expressed
in the coordinates $f^2$,..,$f^n$,$\phi ^1$,..,$\phi ^n$, while
$\Xi :=\mathrm{Re}\,\omega ^M$ and $\Upsilon :=\mathrm{Im}\,\omega ^M$ will
be real special Killing forms in the classical sense.
\end{remark}

\section{An application: the spaces $Y^{p,q}$}

In the particular framework represented by the spaces $Y^{p,q}$, the special
Killing forms were extracted by a direct calculation \cite{Vis}. In this
final section we present an alternative approach using the above results in
this particular case.

Let us consider  the explicit local metric of the 5-dimensional
$Y^{p,q}$ manifold given by the line element \cite{M-S}
\begin{equation}
\begin{split}\label{Ypq}
ds^2 & = \frac{1-c\, y}{6}( d \theta^2 + \sin^2 \theta\, d \phi^2)
+  \frac{1}{w(y)q(y)} dy^2
+ \frac{q(y)}{9} ( d\psi - \cos \theta \, d \phi)^2\\
& \quad  +
w(y)\left[ d\alpha + \frac{ac -2y+ c\, y^2}{6(a-y^2)}
[d\psi - \cos\theta \, d\phi]\right]^2\,,
\end{split}
\end{equation}
where
\begin{equation}
\begin{split}
w(y) & = \frac{2(a-y^2)}{1-cy}\,,\\
q(y) & = \frac{a-3y^2 + 2c y^3}{a-y^2}\,.
\end{split}
\end{equation}

This metric is Einstein with $Ric_g = 4 g$ for all values of the
constants $a,c$. Moreover, the space is also Sasaki. For $c=0$, the
metric takes the local form of the standard homogeneous metric on
$T^{1,1}$ \cite{M-S}. Otherwise, the constant $c$ can be rescaled by a
diffeomorphism, so in what follows we take $c=1$. For
\[
0 < \alpha < 1
\]
we can take the range of the angular coordinates $(\theta, \phi, \psi)$
to be $0\leq \theta \leq 2\pi\,, 0\leq \phi \leq 2\pi\,, 0\leq \psi \leq
2\pi$. Choosing $0 < a < 1$,  the roots $y_i$  of the cubic equation
\[
a - 3 y^2 + 2 y^3 = 0 \,,
\]
are real, one negative $(y_1)$ and two positive $(y_2, y_3)$. If the
smallest of the positive roots is $y_2$, one can take the range of the
coordinate $y$ to be
\[
y_1\leq y \leq y_2 \,.
\]

Following \cite{M-S, M-S-Y}, for this particular space we take the
complex coordinates
\begin{eqnarray}
z^1 &:=&\log \left(r^3\sin \theta \sqrt{\frac{p(y)(1-y)}2}e^{\mathrm{i}\psi }
\right) \label{Y}\\
\  &=&3\ln r+\ln \sin \theta +\frac 12\ln \frac{p(y)(1-y)}2+\mathrm{i}
\psi ^{\prime } \,, \nonumber \\
z^2 &:=&\frac 1{3\sqrt{3}}\log \left( \tan \frac \theta 2e^{\mathrm{i}\phi }
\right) \nonumber \\
\  &=&\frac 1{3\sqrt{3}}\ln \tan \frac \theta 2+\mathrm{i}\phi ^{\prime }
\,, \nonumber\\
z^3 &:=&\frac 16\log \left( \frac 1{\sin \theta }\sqrt{\left( y-y_1\right)
^{-\frac 1{y^1}}\left( y_2-y\right) ^{-\frac 1{y^2}}\left( y_3-y\right)
^{-\frac 1{y^3}}}e^{-6\mathrm{i}\alpha -\mathrm{i}\psi }\right)   \nonumber \\
\  &=&-\frac 16\ln \sin \theta -\frac 12\ln \left( \left( y-y_1\right)
^{-\frac 1{y^1}}\left( y_2-y\right) ^{-\frac 1{y^2}}\left( y_3-y\right)
^{-\frac 1{y^3}}\right)   \nonumber \\
&&\ +\mathrm{i}\beta ^{\prime } \,,  \nonumber
\end{eqnarray}
where we denote (see also \cite{Vis})
\begin{eqnarray}
p(y) &:=& w (y) \cdot q(y) =  \frac{2\left( a-3y^2+2y^3\right)}{a-y^2}\,,
\label{prime}\\
\beta ^{\prime } &:=& -\alpha - \frac{1}{6}\psi \,, \nonumber   \\
\phi ^{\prime } &:=&\frac 1{3\sqrt{3}}\phi \,,  \nonumber \\
\psi ^{\prime } &:=&\psi \,.  \nonumber
\end{eqnarray}
Consequently
\begin{eqnarray}
d\beta ^{\prime } &=&-d\alpha -\frac 16d\psi \,,  \label{d_prime} \\
d\phi ^{\prime } &=&\frac 1{3\sqrt{3}}d\phi \,,  \nonumber\\
d\psi ^{\prime } &=&d\psi\,. \nonumber
\end{eqnarray}

From (\ref{Y}), we obtain
\begin{eqnarray}
x^1 &=&3\ln r+\ln \sin \theta +\frac 12\ln \frac{p(y)(1-y)}2 \,, \label{X} \\
x^2 &=&\frac 1{3\sqrt{3}}\ln \tan \frac \theta 2 \,,  \nonumber \\
x^3 &=&-\frac 16\ln \sin \theta -\frac 12\ln \left( \left( y-y_1\right)
^{-\frac 1{y^1}}\left( y_2-y\right) ^{-\frac 1{y^2}}\left( y_3-y\right)
^{-\frac 1{y^3}}\right) \,,  \nonumber
\end{eqnarray}
while for the toric coordinates we simply have
\begin{eqnarray}
\phi ^1 &=&\psi ^{\prime } \,,  \label{Phi} \\
\phi ^2 &=&\phi ^{\prime } \,,  \nonumber \\
\phi ^3 &=&\beta ^{\prime } \,.  \nonumber
\end{eqnarray}

We derive from (\ref{Y}) the following useful relations (see also \cite{M-S}).
\begin{eqnarray*}
dz^1 &=&\frac{\cos \theta }{\sin \theta }d\theta +3\frac{dr}r-6\frac
y{p(y)}dy+\mathrm{i}d\psi ^{\prime } \,, \\
dz^2 &=&\frac 1{3\sqrt{3}}\frac 1{\sin \theta }d\theta +\mathrm{i}d\phi
^{\prime } \,, \\
dz^3 &=&-\frac 16\frac{\cos \theta }{\sin \theta }d\theta +\frac
1{p(y)}dy+\mathrm{i}d\beta ^{\prime } \,.
\end{eqnarray*}

Now, comparing for instance with \cite[(2.28)]{M-S}, it is easy to verify that
\[
\Omega =e^{z^1}dz^1\wedge dz^2\wedge dz^3 \,.
\]
The above results, obtained in the general case of the metric cone of a
Sasaki-Einstein manifold, can be now applied to calculate the special Killing
forms in our particular case.

From the relation (\ref{X}) and (\ref{Phi}) it follows that we can consider
the foliated coordinates $\left( r;\theta ,y,\psi^{\prime } ,\phi ^{\prime },\beta^{\prime }\right) $; with respect to these convenient local coordinates, the Jacobi matrix $\mathcal{A}$ can be calculated as follows

\[
\mathcal{A}=\left(
\begin{array}{ccc}
\frac 3r & \frac{\cos \theta }{\sin \theta } & -\frac{6y}{p(y)} \\
0 & \frac 1{3\sqrt{3}}\frac 1{\sin \theta } & 0 \\
0 & -\frac 16\frac{\cos \theta }{\sin \theta } & \frac 1{p(y)}
\end{array}
\right) \,.
\]

With the above consideration, we can now proceed to calculate the complex
differential form $\omega ^M$. First of all we have $\frac{\partial x^1}
{\partial r}\not =0$, all other terms of this type vanish, so the only
possibility for $i$ is $1$; consequently $p=0$, there is no $j_s$, and $q$
can only be $0$, $1$ or $2$.

Using the relation (\ref{suma}), we get
\begin{eqnarray}
\omega ^M &=&\frac 1{r^2}\frac 3r\left( r^3\sqrt{\frac{p\left( y\right)
\left( 1-y\right) }2}\sin \theta \right) e^{\mathrm{i}\psi ^{\prime }}
\label{suma1} \\
&&\ \times \left( \left( -1\right) ^0\mathrm{i}^0\mathcal{A}_{0;1;2}^{23}d\theta
\wedge dy+\left( -1\right) ^1\mathrm{i}^1\mathcal{A}_{0;1;2}^2d\theta \wedge
d\phi^{\prime }\right.  \nonumber \\
&&\ +\left( -1\right) ^1\mathrm{i}^1\mathcal{A}_{0;1;2}^3dy\wedge d\phi ^{\prime}
+\left( -1\right) ^0\mathrm{i}^1\mathcal{A}_{0;1;3}^2d\theta \wedge d\beta ^{\prime }
\nonumber \\
&&\ \left. \left( -1\right) ^0\mathrm{i}^1\mathcal{A}_{0;1;3}^3dy\wedge d\beta^{\prime }
+\left( -1\right) ^0\mathrm{i}^2\mathcal{A}_{0;1;2,3}^0d\phi ^{\prime}\wedge
d\beta ^{\prime }\right) \,.  \nonumber
\end{eqnarray}

We calculate the coefficients in the following.

For $q=0$ we obtain
\begin{eqnarray*}
\mathcal{A}_{0;1;2}^{23} &=&\left|
\begin{array}{cc}
\frac 1{3\sqrt{3}}\frac 1{\sin \theta } & 0 \\
-\frac 16\frac{\cos \theta }{\sin \theta } & \frac 1{p(y)}
\end{array}
\right| \\
&=&\frac 1{3\sqrt{3}}\frac 1{\sin \theta p(y)} \,.
\end{eqnarray*}

Next, for $q=1$ one can calculate
\[
\begin{tabular}{ll}
$\mathcal{A}_{0;1;2}^2=-\frac 16\frac{\cos \theta }{\sin \theta },
$ & $\mathcal{A}_{0;1;2}^3=\frac 1{p(y)},$ \\
$\mathcal{A}_{0;1;3}^2=\frac 1{3\sqrt{3}}\frac 1{\sin \theta },
$ & $\mathcal{A}_{0;1;3}^3=0$ \,.
\end{tabular}
\]

Finally, for $q=2$ we obtain
\[
\mathcal{A}_{0;1;2,3}^0=1 \,.
\]
Plugging now all the above terms in the formula (\ref{suma1}), we end up with

\begin{eqnarray*}
\omega ^M &=&e^{\mathrm{i}\psi ^{\prime }}\sqrt{\frac{1-y}{6p\left( y\right)}}
\biggl(d\theta \wedge dy+\mathrm{i}\frac{\sqrt{3}}2\cos \theta p(y)d\theta
\wedge d\phi^{\prime }\biggr. \\
&&\ -\mathrm{i}3\sqrt{3}\sin \theta dy\wedge d\phi ^{\prime }+
\mathrm{i}p(y)d\theta \wedge
d\beta ^{\prime } \\
&&\ \biggl. -3\sqrt{3}p(y)\sin \theta d\phi ^{\prime }\wedge d\beta ^{\prime}\biggr)\,.
\end{eqnarray*}
We write $\omega ^M$ in a more convenient form,
\begin{eqnarray*}
\omega ^M &=&\sqrt{\frac{1-y}{6p\left( y\right) }}\left( \cos \psi ^{\prime}
+\mathrm{i}\sin \psi ^{\prime }\right) \times \Biggl( \left[ -dy\wedge d\theta
+3\sqrt{3}p(y)\sin \theta d\beta ^{\prime }\wedge d\phi ^{\prime }\right]
\Biggr. \\
&&\ \Biggl. +\mathrm{i}\biggl[ -p(y)d\beta ^{\prime }\wedge d\theta +
\frac{\sqrt{3}}
2\cos \theta p(y)d\theta \wedge d\phi ^{\prime }-3\sqrt{3}\sin \theta
dy\wedge d\phi ^{\prime }\biggr] \Biggr) \,.
\end{eqnarray*}

From here, calculating the real and imaginary part, we get
\begin{eqnarray*}
\Xi &=&\sqrt{\frac{1-y}{6p\left( y\right) }}\Biggl( \cos \psi ^{\prime
}\left[ -dy\wedge d\theta +3\sqrt{3}p(y)\sin \theta d\beta ^{\prime }\wedge
d\phi ^{\prime }\right] \Biggr. \\
&&\ -\sin \psi ^{\prime }\biggl[ -3\sqrt{3}\sin \theta dy\wedge d\phi
^{\prime }-p(y)d\beta ^{\prime }\wedge d\theta \biggr. \\
&&\ \biggl. +\Biggl. \frac{\sqrt{3}}2\cos \theta p(y)d\theta \wedge d\phi
^{\prime }\biggr] \Biggr) \,,
\end{eqnarray*}
and respectively
\begin{eqnarray*}
\Upsilon &=&\sqrt{\frac{1-y}{6p\left( y\right) }}\Biggl( \cos \psi ^{\prime
}\biggl[ -3\sqrt{3}\sin \theta dy\wedge d\phi ^{\prime }-p(y)d\beta ^{\prime
}\wedge d\theta \biggr. \Biggr. \\
&&\ \biggl. +\frac{\sqrt{3}}2\cos \theta p(y)d\theta \wedge d\phi ^{\prime
}\biggr] +\sin \psi ^{\prime }\biggr[ -dy\wedge d\theta \biggr. \\
&&\ \Biggl. \biggr. +3\sqrt{3}p(y)\sin \theta d\beta ^{\prime }\wedge d\phi
^{\prime }\biggr] \Biggr) \,.
\end{eqnarray*}

Replacing now $\psi ^{\prime }$, $\phi ^{\prime }$, $\beta ^{\prime }$ by
$\psi $, $\phi $, $\beta $ and considering also (\ref{prime}) and (\ref
{d_prime}), we eventually obtain (14) and (15) from \cite{Vis}.

\section{Conclusions}

It is well-known that a lot of examples of Sasaki-Einstein manifolds may
be obtained via toric geometry, and such examples are a good testing ground
for the predictions of the AdS/CFT correspondence \cite{M-S-Y}. In this
paper we give a general scheme to construct the special Killing forms on a
toric Sasaki-Einstein manifold. This procedure is effectively exemplified
in the case of the $Y^{p,q}$ manifolds, and computations in this explicit
$5-$dimensional space reveal that the special Killing forms obtained in
this article agree with the results previously obtained in \cite{Vis} with
a different approach.

On the other hand, in \cite{MS2} the authors investigated two explicit
infinite families $Y^{p,q}$ of Sasaki-Einstein 7-manifolds, which are
lens-space bundles $S^3/\mathbb{Z}_p$ over $CP^2$ and $CP^1\times CP^1$,
respectively, and showed that the metric cones over these Sasaki-Einstein
$7-$manifolds are in fact toric. It is the hope of the authors that the
techniques developed in this paper can be applied to extract the special
Killing forms on the 7-dimensional $Y^{p,q}$ spaces,
as well on the $L^{abc}$ spaces. We note that the spaces $L^{abc}$
(where $a,b,c$ are three positive coprime integers that satisfy some other
relations), are certain 5-dimensional Einstein-Sasaki manifolds with a
$T^3$-worth of isometries acting with cohomogeneity two \cite{CC}. These
metrics are constructed starting by rotating anti-de Sitter black
hole metrics in 5 dimensions, Euclideanizing, and then taking an
appropriate limit motivated by physical considerations. The restrictions
on $a,b,c$ arise from requiring that the locally defined metrics extend
smoothly over a compact, non-singular 5-manifold.  When $a+b=2c$, these
metrics reduce to the cohomogeneity one metrics $Y^{p,q}$.

\section*{Acknowledgments}
The authors would like to thank the referee for his valuable comments and suggestions
which helped to improve the paper. M. Visinescu was supported by CNCS-UEFISCDI, project number
PN-II-ID-PCE-2011-3-0137. The work of
G.E. V\^{\i}lcu  was supported by CNCS-UEFISCDI, project number
PN-II-ID-PCE-2011-3-0118.

\end{document}